\documentclass[aps,prd,twocolumn,groupedaddress,showpacs,amsmath,amssymb,nofootinbib]{revtex4}

\usepackage{graphicx}
\usepackage{bm}
\usepackage{amsmath}

\begin{document}


\title{Probing 
pretzelosity $h^{\perp}_{1T}$ via the polarized proton-antiproton
Drell-Yan process}



\newcommand*{\PKU}{School of Physics and State Key Laboratory
of Nuclear Physics and Technology, \\Peking University, Beijing
100871, China}\affiliation{\PKU}
\newcommand*{\CHEP}{Center for High Energy
Physics, Peking University, Beijing 100871,
China}\affiliation{\CHEP}

\author{Jiacai Zhu}\affiliation{\PKU}
\author{Bo-Qiang Ma}\email{mabq@pku.edu.cn}\affiliation{\PKU}\affiliation{\CHEP}



\begin{abstract}
We show that the polarized proton-antiproton Drell-Yan process is
ideal to probe the pretzelosity distribution ($h_{1T}^\perp$), which
is one of the new transverse-momentum-dependent parton distributions
of the nucleon. We present predictions of the $\cos(2\phi + \phi_a -
\phi_b)$ asymmetry in the transversely polarized proton-antiproton
Drell-Yan process at PAX kinematics and find that the results are
significantly larger compared with those of the $\sin(3\phi_h -
\phi_S)$ asymmetry in the semi-inclusive deep inelastic scattering
process at HERMES, COMPASS, and JLab kinematics. We conclude that the
$\cos(2\phi + \phi_a - \phi_b)$ asymmetry in the PAX experiment can
probe the new physical quantity of the pretzelosity distribution.
\end{abstract}

\pacs{13.85.Qk, 13.88.+e, 14.20.Dh, 25.43.+t}

\maketitle

At leading-twist level, the quark structure of the nucleon can be
described by three parton distribution functions: the
longitudinal momentum distribution $f_1(x)$, the helicity
distribution $g_1(x)$, and the transversity distribution $h_1(x)$.
The first two of them have been determined with high precision from
the inclusive deep inelastic scattering and Drell-Yan
experiments. The transversity distribution is chiral-odd and
thus is not observable in the inclusive deep inelastic scattering by chirality conservation
but can be probed when it is combined with another
chiral-odd function, such as itself through the Drell-Yan
process. The transversity distribution can be also accessed with the
combination of the Collins function
through the semi-inclusive deep inelastic scattering process, in which the
intrinsic quark transversal motions are involved. Thus the transversity distribution,
as well as the unpolarized distribution and the
helicity distribution,
needs to be extended
to the transverse-momentum-dependent parton distributions (TMDs) as
a generalization of parton distribution functions from one dimension to three
dimensions in momentum space. At leading twist, there are eight
TMDs including five new additional ones; among them,
the Sivers distribution $f_{1T}^\perp(x,k_T^2)$ and its chiral-odd
partner, the Boer-Mulders distribution $h_1^\perp(x,k_T^2)$, are
well known for their T-odd property---i.e., they change sign under
naive time reversal. Recently, the pretzelosity distribution has been
under attention widely~\cite{Avakian2008,Pasquini2008, Ma2009,
Efremov2009, Avakian2010,Bacchetta2008,Boffi2009}.

The pretzelosity distribution ($h_{1T}^\perp$) is one of the eight
leading-twist 
TMDs.
In a bag model calculation~\cite{Avakian2008}, the
pretzelosity distribution has an interesting relation with the
helicity and transversity distributions:
\begin{equation}
h_{1T}^{\perp(1)q} (x, k_T^2) \equiv   \frac{k_T^2}{2M_N^2}
h_{1T}^{\perp q}(x, k_T^2)  =   g_1^q(x, k_T^2) - h_1^q(x, k_T^2).
\end{equation}
This relation is also valid in several other quark model
calculations~\cite{Pasquini2008, Ma2009, Efremov2009, Avakian2010}.
In a spectator model~\cite{Bacchetta2008}, the relation is not
satisfied for the axial-vector coupling case, but it remains valid
for the scalar coupling case. Moreover, in a light-cone
quark-diquark model~\cite{Ma2009}, it is found that the pretzelosity
distribution is closely connected with the quark orbital angular
momentum distribution
\begin{equation}
L^{qv} (x, k_T^2) = - h_{1T}^{\perp(1)qv} (x, k_T^2),
\end{equation}
and such a relation is also supported in the bag
model~\cite{Avakian2010}. Though these results might be model
dependent, the pretzelosity is surely a new physical quantity that can
provide important information on the quark spin and orbital
correlation of the nucleon.

The pretzelosity distribution is chiral-odd, so similar to
the transversity, it could be measured through the $\sin(3\phi_h -
\phi_S)$ asymmetry in the semi-inclusive deep inelastic scattering
when combined with the Collins function.
The predictions of the asymmetry in Refs.~\cite{Ma2009, Boffi2009}
turn out to be rather small, up to $1 \%$. If one makes a transverse
momentum cut in data analysis~\cite{Ma2009}, the results could be
enhanced up to $4 \%$.

When one combines two chiral-odd TMDs, for example, the transversity
and pretzelosity distributions in the polarized Drell-Yan process, it is
also possible to access the pretzelosity distribution. But the
double spin asymmetry $A_{TT}$ in the proton-proton Drell-Yan process is
usually small because $A_{TT}$ is proportional to the production of
the polarized quark distribution with the antiquark one. However,
the situation is different in polarized proton and antiproton
collisions, where the double spin asymmetry $A_{TT}$ is proportional
to the production of the polarized quark distribution from the
proton and the polarized antiquark distribution from the antiproton.
Such an experiment has been proposed by the polarized
antiproton experiment (PAX)
Collaboration~\cite{pax2005}, and there has been new technical
progress~\cite{pax2010} toward the goal for a proton-antiproton
collider with both beams polarized~\cite{Rathmann:2004pm}. The PAX
experiment was originally proposed to access the transversity
distributions; as the transversity distributions have been probed in
other experiments, such as HERMES~\cite{hermes2005, hermes2010},
COMPASS~\cite{compass2005, compass2010}, and JLab~\cite{jlab, Gao},
probing the pretzelosity distributions could be a new motivation for
the PAX experiment. The purpose of this paper is to show that the
polarized proton-antiproton Drell-Yan process is ideal to measure
the pretzelosity distributions of the nucleon.

The leading order differential cross section for the double
transversely polarized Drell-Yan process reads~\cite{drell-yan}
\begin{widetext}
\begin{equation}
  \frac{d\sigma}{dx_a dx_b d\bm{q}_T d\Omega}
=  \frac{\alpha^2_{em}}{4 Q^2} \Big\{ F_{UU}^1 +
\lvert\bm{S}_{aT}\rvert \lvert \bm{S}_{bT} \rvert \sin^2\theta
    \big[\cos(2\phi - \phi_a - \phi_b) F_{TT}^{\cos(2\phi - \phi_a -
 \phi_b)}
  + \cos(2\phi + \phi_a - \phi_b) F_{TT}^{\cos(2\phi + \phi_a - \phi_b)} \big] + \ldots \Big\}.
\label{eq:drell-yan}
\end{equation}
The subscripts $a$ and $b$ stand for the incoming hadrons in the Drell-Yan
process, and $\phi_a$ and $\phi_b$ are the angles of $\bm{S}_{aT}$
and $\bm{S}_{bT}$, respectively. The structure functions in
Eq.~(\ref{eq:drell-yan}) are
\begin{equation}
F_{UU}^1  =  \mathcal{C} \big[ f_1\bar{f}_1 \big],  ~~~~
F_{TT}^{\cos(2\phi - \phi_a - \phi_b)}  = \mathcal{C} \big[
h_1\bar{h}_1 \big], ~~~~ F_{TT}^{\cos(2\phi + \phi_a - \phi_b)}  =
\mathcal{C} \Big[ \frac{2(\bm{h}\cdot\bm{k}_{aT})^2 -
k_{aT}^2}{2M_a^2} h_{1T}^\perp\bar{h}_1 \Big]\label{eq:pretz_struc},
\end{equation}
where the notations $\bm{h}=\bm{q}_T/q_T$ and
\begin{equation}
\begin{split}
 \mathcal{C} \big[ w(\bm{k}_{aT}, \bm{k}_{bT}) f_1 \bar{f}_2 \big]
\equiv  \frac{1}{N_c} \sum_q e_q^2 \int d\bm{k}_{aT} d\bm{k}_{bT} \,
\delta^{(2)}(\bm{q}_T - \bm{k}_{aT} - \bm{k}_{bT})
    w(\bm{k}_{aT}, \bm{k}_{bT}) &\big[ f_1^q(x_a, k_{aT}^2) f_2^{\bar{q}}(x_b,
 k_{bT}^2)
 \\
  & + f_1^{\bar{q}}(x_a, k_{aT}^2)
f_2^q(x_b, k_{bT}^2)\big] \end{split}
\end{equation}
\end{widetext}
are used. Other terms will not contribute in our analysis below, but
we should mention that another structure function
$F_{TT}^{\cos(2\phi - \phi_a + \phi_b)}$ in Ref.~\cite{drell-yan}
involves the convolution of $h_1$ and $\bar{h}_{1T}^\perp$ and can
be obtained from $F_{TT}^{\cos(2\phi + \phi_a - \phi_b)}$ by switching
the subscript labels $a$ and $b$. Then we obtain the
$\cos(2\phi - \phi_a - \phi_b)$ and
$\cos(2\phi + \phi_a - \phi_b)$ asymmetries
\begin{align}
A_{TT}^{\cos(2\phi - \phi_a - \phi_b)} & = \frac{\frac{\alpha^2_{em}}{4 Q^2} F_{TT}^{\cos(2\phi - \phi_a - \phi_b)}}{\frac{\alpha^2_{em}}{4 Q^2} F_{UU}^1}, \\
A_{TT}^{\cos(2\phi + \phi_a - \phi_b)} & = \frac{\frac{\alpha^2_{em}}{4 Q^2} F_{TT}^{\cos(2\phi + \phi_a - \phi_b)}}{\frac{\alpha^2_{em}}{4 Q^2} F_{UU}^1}.
\end{align}

Considering polarized proton and antiproton collisions, with charge conjugation invariance
\begin{align}
f_{\bar{p}}^q(x, k_T^2) & = f_p^{\bar{q}}(x, k_T^2), \\
f_{\bar{p}}^{\bar{q}}(x, k_T^2) & = f_p^q(x, k_T^2),
\end{align}
we get
\begin{align}
A_{TT}^{\cos(2\phi - \phi_a - \phi_b)} & = \frac{\frac{\alpha^2_{em}}{4 Q^2} \mathcal{C} \big[ h_1 h_1 \big]}{\frac{\alpha^2_{em}}{4 Q^2}\mathcal{C} \big[ f_1 f_1 \big]}, \\
A_{TT}^{\cos(2\phi + \phi_a - \phi_b)} & = \frac{\frac{\alpha^2_{em}}{4 Q^2} \mathcal{C} \Big[ \frac{2(\bm{h}\cdot\bm{k}_{aT})^2 - k_{aT}^2}{2M_N^2} h_{1T}^\perp h_1 \Big]}{\frac{\alpha^2_{em}}{4 Q^2}\mathcal{C} \big[ f_1 f_1 \big]},
\end{align}
and $M_a = M_b = M_N$.
If we fix $Q^2$ and integrate the structure functions upon $\bm{q}_T$, then
\begin{equation}
\begin{split}
& A_{TT}^{\cos(2\phi - \phi_a - \phi_b)}(x_F)\\
=&  \frac{\sum_q e_q^2 [h_1^q(x_a) h_1^q(x_b) +
h_1^{\bar{q}}(x_a)h_1^{\bar{q}}(x_b)]}{\sum_q e_q^2 [f_1^q(x_a)
f_1^q(x_b) + f_1^{\bar{q}}(x_a)f_1^{\bar{q}}(x_b)] },
\end{split}
\label{eq:trans_asy}
\end{equation}
where $x_a$ and $x_b$ are given by
\begin{equation}
x_F = x_a - x_b,~x_ax_b = \frac{Q^2}{s}.
\end{equation}
Using the method introduced in Refs.~\cite{Kotzinian, Ma2007}, we
obtain the weighted asymmetry
\begin{equation}
\begin{split}
  & A_{TT}^{\frac{q_T^2}{2M_N^2}\cos(2\phi + \phi_a - \phi_b)}(x_F) \\
= & \frac{\sum_q e_q^2 [h_{1T}^{\perp(2)q}(x_a) h_1^q(x_b) + h_{1T}^{\perp(2)\bar{q}}(x_a)h_1^{\bar{q}}(x_b)]}{\sum_q e_q^2 [f_1^q(x_a) f_1^q(x_b) + f_1^{\bar{q}}(x_a)f_1^{\bar{q}}(x_b)]},
\end{split}
\label{eq:weighted_pretz_asy}
\end{equation}
with
\begin{equation}
h_{1T}^{\perp(2)} (x) \equiv \int d\bm{k}_T
\Big(\frac{k_T^2}{2M_N^2}\Big)^2 h_{1T}^\perp (x, k_T^2).
\end{equation}
For Eq.~(\ref{eq:pretz_struc}), we find
\begin{widetext}
\begin{equation}
\int d \bm{q}_T F_{TT}^{\cos(2\phi + \phi_a - \phi_b)} =  \frac{1}{4
N_c} \sum_q e_q^2 \int d k_{aT}^2 d k_{bT}^2
[ h_{1T}^{\perp q}( x_a ,k_{aT}^2) h_1^q( x_b , k_{bT}^2) + h_{1T}^{\perp\bar{q}}( x_a ,k_{aT}^2) h_1^{\bar{q}}( x_b , k_{bT}^2)]
\times A,
\end{equation}
where
\begin{equation}
\begin{split}
A =   \int d \phi_{aT} d \phi_{bT} \frac{2(\bm{h}\cdot\bm{k}_{aT})^2
- k_{aT}^2 }{2M_N^2}\bigg\lvert_{\bm{q}_T = \bm{k}_{aT} +
\bm{k}_{bT}}
  =   \begin{cases}
0 & k_{aT} \leqslant k_{bT}, \\
\frac{k_{aT}^2 - k_{bT}^2}{2M_a^2} \int d \phi_{aT} ~ d \phi_{bT} &
k_{aT} > k_{bT}.
\end{cases}
\end{split}
\end{equation}
\end{widetext}

Therefore
\begin{equation}
\int d \bm{q}_T F_{TT}^{\cos(2\phi + \phi_a - \phi_b)} \not\equiv 0,
\end{equation}
and the $\cos(2\phi + \phi_a - \phi_b)$ asymmetry
\begin{equation}
\begin{split}
  & A_{TT}^{\cos(2\phi + \phi_a - \phi_b)}(x_F)\\
= & \frac{N_c \int d \bm{q}_T F_{TT}^{\cos(2\phi + \phi_a -
\phi_b)}(x_a, x_b, \bm{q}_T)}{\sum_q e_q^2 [f_1^q(x_a) f_1^q(x_b) +
f_1^{\bar{q}}(x_a)f_1^{\bar {q}}(x_b)]} \not\equiv 0.
\end{split}
\label{eq:pretz_asy}
\end{equation}
The double spin asymmetry $A_{TT}^{\cos(2\phi - \phi_a - \phi_b)}$
in Eq.~(\ref{eq:trans_asy}) reflects pure information on the
transversity distribution, whereas
$A_{TT}^{\frac{q_T^2}{2M_N^2}\cos(2\phi + \phi_a - \phi_b)}$ in
Eq.~(\ref{eq:weighted_pretz_asy}) and $A_{TT}^{\cos(2\phi + \phi_a -
\phi_b)}$ in Eq.~(\ref{eq:pretz_asy}) are new asymmetries involving
both transversity and pretzelosity distributions.
Thus these asymmetries provide us the possibility to pin down the
pretzelosity distribution, and we present predictions of them below.

In the light-cone quark-diquark model~\cite{Ma2000}, the
relativistic effect due to the quark transversal motions through the
Melosh-Wigner rotation~\cite{Ma91} is important to understand the
helicity distribution suppression compared to the naive quark model
expectation~\cite{Ma:2001ui}. The pretzelosity distributions for up
and down valence quarks are~\cite{Ma2009}
\begin{align}
h^{\perp(uv)}_{1T}(x, k_T^2) &= -\frac{1}{16\pi^3} 
(\frac{1}{9} \sin^2\theta_0 \varphi_V^2 W_V
-\cos^2\theta_0 \varphi_S^2 W_S), \nonumber\\
h^{\perp(dv)}_{1T}(x, k_T^2) &= -\frac{1}{8\pi^3} 
\frac{1}{9} \sin^2\theta_0 \varphi_V^2 W_V,
\label{eq:pretz1}
\end{align}
with the Melosh-Wigner rotation factor $W_D (D = V,
S)$~\cite{Ma:1998ar}
\begin{equation}
W_D(x, k_T^2) = -\frac{2M_N^2}{(x\mathcal{M}_D + m_q)^2 + k_T^2},
\end{equation}
where $\mathcal{M}_D = \sqrt{(m_q^2 + k_T^2)/x + (m_D^2 + k_T^2)/(1
- x)}$. The transversity distributions for up and down valence
quarks have similar forms to Eqs.~(\ref{eq:pretz1}), but with
different Melosh-Wigner rotation factor $\tilde{W}_D (D = V,
S)$~\cite{Ma1998}
\begin{equation}
\tilde{W}_D(x, k_T^2) = \frac{(x\mathcal{M}_D + m_q)^2
}{(x\mathcal{M}_D + m_q)^2 + k_T^2}.
\end{equation}
The unpolarized distributions can be found in Ref.~\cite{Ma2000}
\begin{align}
f_1^{(uv)}(x, k_T^2) &= \frac{1}{16\pi^3}   (\frac{1}{3} \sin^2\theta_0 \varphi_V^2 + \cos^2\theta_0 \varphi_S^2),\nonumber\\
f_1^{(dv)}(x, k_T^2) &= \frac{1}{8\pi^3}  \frac{1}{3} \sin^2\theta_0
\varphi_V^2. \label{eq:unpol}
\end{align}
$\varphi_D (D = V, S)$ is the wave function in the momentum space
for the quark-diquark, and for which we can use the
Brodsky-Huang-Lepage prescription~\cite{Brodsky82,Huang1994}:
\begin{equation}
\varphi_D(x, k_T^2)=A_D\exp\Big\{-\frac{1}{8\alpha_D^2}\big[\frac{m_q^2+k_T^2}{x}+\frac{m_D^2+k_T^2}{1-x}\big]\Big\}.
\end{equation}
The pretzelosity distributions can be expressed with the unpolarized distributions by using
Eqs.~(\ref{eq:pretz1}) and Eqs.~(\ref{eq:unpol})
\begin{align}
h^{\perp(uv)}_{1T}(x, k_T^2) =& \big[f_1^{(uv)}(x, k_T^2) - \frac{1}{2} f_1^{(dv)}(x, k_T^2)\big] W_S(x, k_T^2) \nonumber\\
&-\frac{1}{6} f_1^{(dv)}(x, k_T^2) W_V(x, k_T^2),\nonumber\\
h^{\perp(dv)}_{1T}(x, k_T^2) =& -\frac{1}{3} f_1^{(dv)}(x, k_T^2) W_V(x, k_T^2),
\label{eq:pretz2}
\end{align}
and the transversity distributions also have similar forms.

\begin{figure}
\includegraphics[width=0.80\columnwidth]{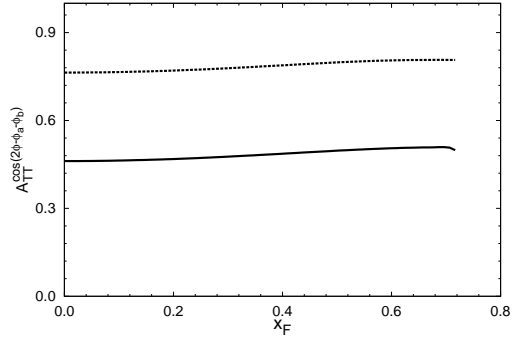}
\caption{\label{fig:trans}The $\cos(2\phi - \phi_a -\phi_b)$
asymmetry as a function of $x_F$ for $s = 45 ~ \mathrm{GeV}^2$ and
$Q^2 = 12 ~ \mathrm{GeV}^2$. The dashed curve corresponds to approach 1,
while the solid curve corresponds to approach 2.}
\end{figure}

We present numerical calculations in two different approaches: For
approach 1, we use Eqs.~(\ref{eq:pretz1}), Eqs.~(\ref{eq:unpol}), and
the transversity version of Eqs.~(\ref{eq:pretz1}) directly to
calculate, while, for approach 2, we adopt the CTEQ6L
parametrization~\cite{cteq} for the unpolarized distributions,
assume a Gaussian form factor of transverse momentum as suggested in
Ref.~\cite{Anselmino2005}:
\begin{equation}
f_1(x, k_T^2) = f_1(x) \frac{\exp(-k_T^2 / k_{av}^2)}{\pi k_{av}^2}
\end{equation}
with $k_{av}^2 = 0.25~\mathrm{GeV}^2$, and then use
Eqs.~(\ref{eq:pretz2}) and their transversity version to calculate.
We should mention that we only sum over the valence quark
distributions in approach 1, while in approach 2 the unpolarized
quark and antiquark distributions are considered. The parameters
that we use in numerical calculations for approach 1 can be found in
Ref.~\cite{Ma2000} as shown in Table.~\ref{table:parameters}, and
$\theta_0$ is fixed to $\pi/4$. According to the discussion in
Ref.~\cite{pax2005}, we choose the kinematics as $s =
45~\mathrm{GeV}^2$ and $Q^2 = 12~\mathrm{GeV}^2$. As we mentioned in
Ref.~\cite{Ma2009}, approach 2 involves CTEQ6L parametrization,
which has been well verified and constrained by many experiments,
and can give more reasonable predictions for future experiments.

\begin{table}[htbp]
\caption{\label{table:parameters}Parameters of the light-cone
quark-diquark model}
\begin{tabular}{cccc}
\hline \hline
$\alpha_D$(GeV) & $m_q$(GeV) & $m_S$(GeV) & $m_V$(GeV) \\
\hline
0.33 & 0.33 & 0.60 & 0.80\\
\hline\hline
\end{tabular}
\end{table}

\begin{figure}
\includegraphics[width=0.80\columnwidth]{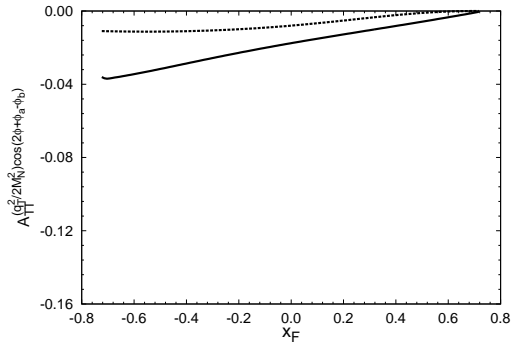}
\caption{\label{fig:weight_pretz}The weighted $\cos(2\phi + \phi_a
-\phi_b)$ asymmetry as a function of $x_F$ for $s = 45 ~
\mathrm{GeV}^2$ and $Q^2 = 12 ~ \mathrm{GeV}^2$. The dashed curve
corresponds to approach 1, while the solid curve corresponds to
approach 2.}
\end{figure}

\begin{figure}
\includegraphics[width=0.80\columnwidth]{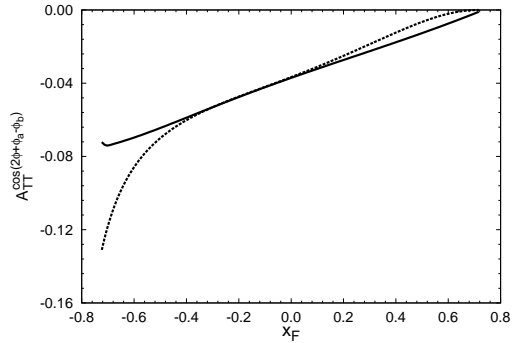}
\caption{\label{fig:pretz}The unweighted $\cos(2\phi + \phi_a
-\phi_b)$ asymmetry as a function of $x_F$ for $s = 45 ~
\mathrm{GeV}^2$ and $Q^2 = 12 ~ \mathrm{GeV}^2$. The dashed curve
corresponds to approach 1, while the solid curve corresponds to
approach 2.}
\end{figure}

The results for the $\cos(2\phi - \phi_a - \phi_b)$ asymmetry in
Eq.~(\ref{eq:trans_asy}) are shown in Fig.~\ref{fig:trans}, and the
magnitude is comparable with the results obtained in
Ref.~\cite{Efremov2004} to measure the transversity. For the
$\cos(2\phi + \phi_a -\phi_b)$ asymmetry, we calculate the weighted
one in Eq.~(\ref{eq:weighted_pretz_asy}) and the unweighted one in
Eq.~(\ref{eq:pretz_asy}), and the results are shown in
Figs.~\ref{fig:weight_pretz} and \ref{fig:pretz}, respectively.
The magnitudes of these asymmetries, especially the unweighted one,
are significantly larger compared with those of the $\sin(3\phi_h -
\phi_S)$ asymmetry in the semi-inclusive deep inelastic scattering
process. It is also remarkable to notice that the sizes for the two
approaches change order from the weighted one to the unweighted one, due to
the different transverse momentum ($k_T^2$) dependence of the two
approaches. Such a feature is essential to discriminate different
predictions of the pretzelosity through one single experiment.

For typical PAX kinematics in fixed target mode ($s =
45~\mathrm{GeV}^2$ and $Q^2 = 12~\mathrm{GeV}^2$),  quarks of the
proton and antiquarks of the antiproton at the large $x$ region
contribute dominantly~\cite{pax2005}. Besides, the PAX experiment is
sensitive to polarized up quark distributions $h_1^u$ and
$h_{1T}^{\perp u}$, and the ratio $h_1^u/f_1^u$ at the large $x$ region
is large as shown in Ref.~\cite{Ma2002}. We thus conclude that the
asymmetries of Eqs.~(\ref{eq:weighted_pretz_asy}) and
(\ref{eq:pretz_asy}) in the PAX experiment are feasible to measure the
new physical quantity of the pretzelosity distribution
($h_{1T}^\perp$).

\begin{acknowledgments}
This work is supported by National Natural Science Foundation of
China (Grants No. 10721063, No. 10975003, and No. 11035003).
\end{acknowledgments}

\end{document}